\documentclass[10pt,twocolumn,showpacs,amsmath,amssymb,floatfix,superscriptaddress]{revtex4-2}

\usepackage[normalem]{ulem}
\usepackage{graphicx}
\usepackage{dcolumn}
\usepackage{bm}
\usepackage{color}
\usepackage{txfonts}
\usepackage{microtype}
\usepackage{hyperref}
\usepackage{easyReview}
\usepackage{comment}
\usepackage{braket}
\setreviewson

\begin{document}
	
	\title{Generation of High-Brilliance  Polarized $\gamma$-Rays \textcolor{black}{via Vacuum Dichroism-assisted Vacuum Birefringence}}
	
	\author{Chong Lv}
	\affiliation{Department of Nuclear Physics, China Institute of Atomic Energy, P. O. Box 275(7), Beijing 102413, China}
	
	\author{Feng Wan}\email{wanfeng@xjtu.edu.cn}
	\affiliation{Ministry of Education Key Laboratory for Nonequilibrium Synthesis and Modulation of Condensed Matter, Shaanxi Province Key Laboratory of Quantum Information and Quantum Optoelectronic Devices, School of Physics, Xi'an Jiaotong University, Xi'an 710049, China}

 \author{Yousef I. Salamin}
\affiliation{Department of Physics, American University of Sharjah, Sharjah, POB 26666 Sharjah,  United Arab Emirates}
	
	\author{Qian Zhao}
	\affiliation{Ministry of Education Key Laboratory for Nonequilibrium Synthesis and Modulation of Condensed Matter, Shaanxi Province Key Laboratory of Quantum Information and Quantum Optoelectronic Devices, School of Physics, Xi'an Jiaotong University, Xi'an 710049, China}
	
	\author{Mamutjan Ababekri}
	\affiliation{Ministry of Education Key Laboratory for Nonequilibrium Synthesis and Modulation of Condensed Matter, Shaanxi Province Key Laboratory of Quantum Information and Quantum Optoelectronic Devices, School of Physics, Xi'an Jiaotong University, Xi'an 710049, China}
	
	\author{Ruirui Xu}
	\affiliation{Department of Nuclear Physics, China Institute of Atomic Energy, P. O. Box 275(7), Beijing 102413, China}
	
	\author{Jian-Xing Li}\email{jianxing@xjtu.edu.cn}
	\affiliation{Ministry of Education Key Laboratory for Nonequilibrium Synthesis and Modulation of Condensed Matter, Shaanxi Province Key Laboratory of Quantum Information and Quantum Optoelectronic Devices, School of Physics, Xi'an Jiaotong University, Xi'an 710049, China}
 	\affiliation{Department of Nuclear Physics, China Institute of Atomic Energy, P. O. Box 275(7), Beijing 102413, China}
	\date{\today}
	
	\begin{abstract}
		\textcolor{black}{We put forward a novel method to generate high-brilliance polarized $\gamma$-photon beams via vacuum dichroism (VD)-assisted vacuum birefringence (VB) effect. We split a linearly polarized (LP) laser pulse into two subpulses with the first one colliding with a dense unpolarized electron beam to generate LP $\gamma$ photons (via nonlinear Compton scattering), which then further collide with the second subpulse and are partially transformed into circularly polarized ones via the VB effect. We find that by manipulating the relative polarization of two subpulses, one can ``purify'' (i.e., enhance) the polarization of the $\gamma$-photon beam via the VD effect. Due to the VD assistance, the VB effect reaches optimal when the relative polarization is nearly $30^\circ$, not the widely used $45^\circ$  in the common VB detection methods. In addition, our method can be used to efficiently confirm the well-known VB effect itself, which has not been directly observed in experiments yet.}
\end{abstract}
	
	\maketitle
	
	It is conjectured that ultraintense, high-brilliance, and highly polarized $\gamma$-photons, generated via laser-based means, will play crucial roles in advancing contemporary research in particle physics \cite{Moortgat2008}, nuclear physics \cite{Uggerhoj2005}, astrophysics \cite{Laurent2011}, as well as in many applications in materials science and medicine \cite{Vetter2018}.
	For example, circularly polarized (CP) $\gamma$ photons can act as powerful probes for investigating a wide range of fundamental processes that involve spin angular momentum, including parity violation \cite{Moeini2013}, vacuum birefringence (VB), elastic photon-photon scattering \cite{Micieli2016}, and photoproduction of mesons \cite{Akbar2017}. 
	Currently, high-energy $\gamma$-photon beams are mainly generated in processes involving synchrotron radiation,  bremsstrahlung, and laser-Compton effect.
	Compton scattering of CP light \cite{Bocquet1997, Nakano1998, Omori2006} or bremsstrahlung employing longitudinally spin-polarized (LSP) electrons \cite{Olsen1959, Abbott2016, Kuraev2010, Sun2022cat} can produce CP $\gamma$ photons. 
	However, the generated beam brilliance is limited due to the small scattering cross-section and low brilliance of the LSP electron beam produced in conventional accelerators.
	
	Rapid progress in ultraintense laser technology has pushed the peak intensity available to laboratory experiments to the level of $I_0\simeq 10^{23}~\mathrm{W/cm^2}$  \cite{Yoon_2019_Achieving, Yoon_2021, ELI-beamlines, ELI_np, SULF, ECELS}. 
 Thus, high-brilliance $\gamma$-photon beams can be generated via nonlinear Compton scattering (NCS) of ultraintense laser off dense electron beams produced via laser-plasma wakefield acceleration (LWFA) \cite{Mackenroth2013, Li2015, King2020}.
	In NCS, the yield of $\gamma$ photons increases with increasing laser intensity, and the dominant factor of the emission polarization gradually shifts from the polarization of the scattering laser to the electron beam due to the multiphoton absorption effect \cite{Ivanov2004, wang2023manipulation}. 
	Meanwhile, an  LSP electron beam is mandatory for generating a highly CP $\gamma$-photon beam via NCS \cite{King2020, Li_2020_Polarized} or bremsstrahlung \cite{Neumcke1966, Likhachev2002}. 
	Also, to attain high-brilliance $\gamma$ photons via NCS, the dense LSP electrons can only be generated via LWFA \cite{Wen2019, Nie2021, Nie2022}.
	However, preparing the pre-polarized plasma and maintaining the polarization degree of the accelerated electron beam are still challenging for LWFA.
	At the same time, the circular polarization degree of such $\gamma$ photons increases linearly with energy, but the yield decreases exponentially;  only very high-energy photons can achieve a high circular polarization degree~\cite{Li_2020_Polarized}.
	Thus, it is still a great challenge to generate a high-brilliance and highly CP $\gamma$-photon beam by a feasible unpolarized electron beam.
	
	The recent advances in laser technologies have also revived theoretical and experimental studies of all-optical strong field quantum electrodynamics (QED) effects \cite{Marklund2006, Piazza2012}, such as radiation-reaction effects \cite{Thomas2012, Blackburn2014, Gonoskov2022, Wistisen2018, Cole2018, Poder2018, GonoskovRMP}, NCS \cite{Mackenroth2013, Blackburn2014, Li2015}, nonlinear Breit-Wheeler process (NBW) \cite{Piazza2016, Ridgers2012}, and all related polarization effects \cite{Sorbo2017, King2020, Li_2020_Polarized, Xue2020, Wan_2020, Seipt_2021,wan2023simulations, Sun2024}.
	Ultraintense lasers also facilitate exploration of the polarization properties of the quantum vacuum, such as VB \cite{Klein1964} and vacuum dichroism (VD) \cite{Bragin_2017, King_2016_Vacuum, Heyl_1997}.
	Conventional detection methods of the VB effect mainly focus on the low-energy probe photons traversing the magnetically-polarized vacuum.
	For instance, Polarization of Vacuum with LASer (PVLAS) \cite{DellaValle2016, Ejlli2020}, Observing VAcuum with Laser (OVAL) \cite{Fan2017}, and Bir\'{e}fringence Magn\'{e}tique du Vide (BMV) \cite{Cadene2014} use low-energy photons as probes, and other theoretical proposals also use high-energy probes \cite{Cantatore_1991_Proposed, Wistisen_2013_Vacuum}.
	However, the detection accuracy is still beyond the attainable signal strength. Petawatt laser-based VB detection proposals, involving X-ray free electron lasers (XFEL)  \cite{Heinzl_2006_observation, Karbstein_2016lby, king_heinzl_2016, Shen_2018, Schlenvoigt_2016, Ahmadiniaz2023} or GeV-scale $\gamma$-ray probes \cite{Bragin_2017, Nakamiya_2017, King_2016_Vacuum}, also attract much  attention.
	For instance, VB can be studied by detecting the reduction of linear polarization of the CP $\gamma$ photons, and vice versa, while VD can be investigated by measuring the change of linear polarization due to $e^+e^-$ pair production \cite{King_2016_Vacuum, Bragin_2017, Borysov2022}. Recent proposals have also suggested alternative methods for detecting VB, such as Coulomb-field assistance \cite{Ahmadiniaz2021}, flying focusing \cite{Jin2023}, and plasma effects \cite{PiazzaEnhancement, wan2022enhanced}. Nevertheless, direct observation of the VB effect in experiments with currently available laser systems is still an open question.
 
	\begin{figure}[!t]
		\centering
		\includegraphics[width=\linewidth]{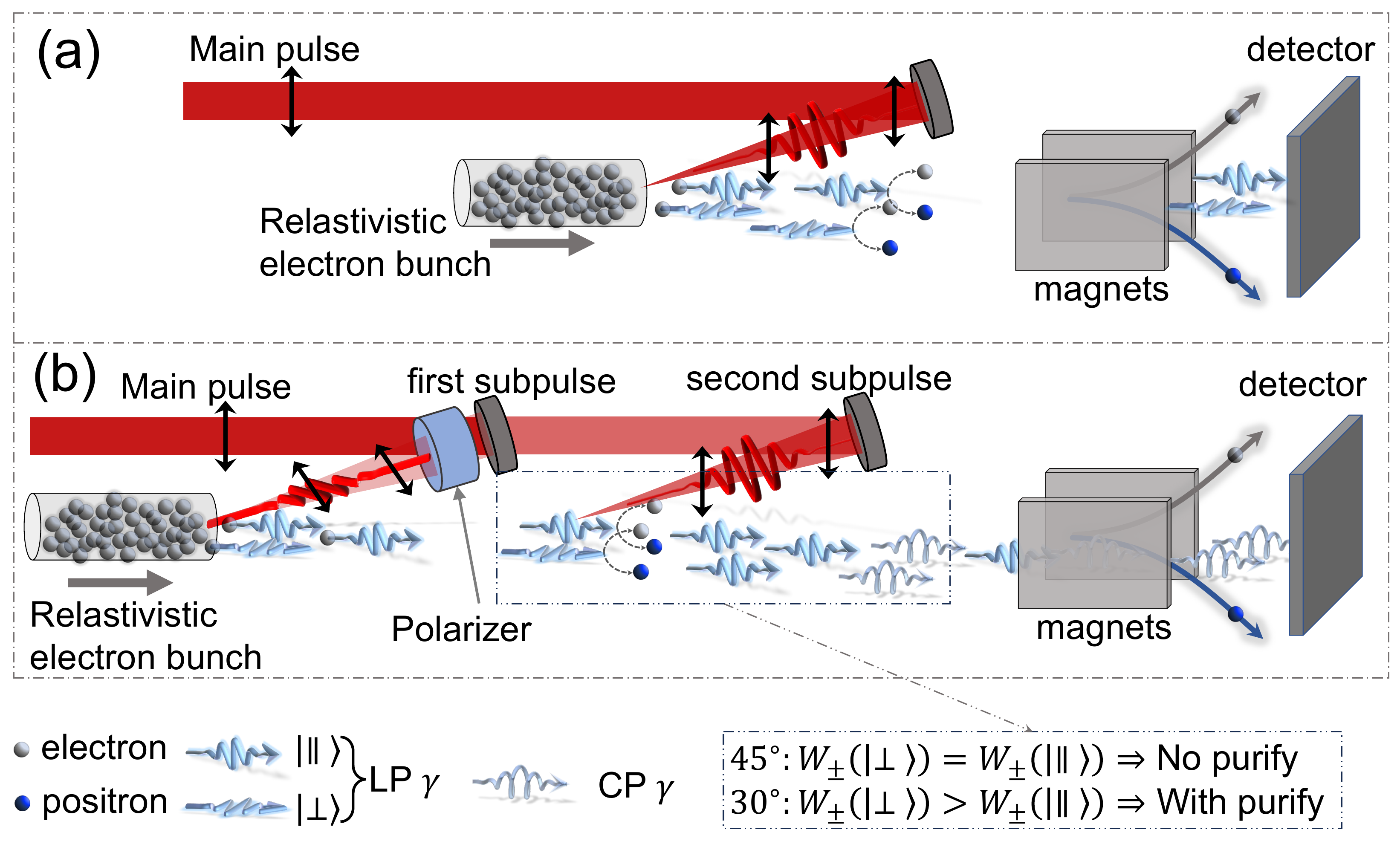}
		\caption{(a) Conventional nonlinear Compton scattering and nonlinear Breit-Wheeler pair production setup. (b) Generation of a CP $\gamma$-photon beam via the VD-assisted VB effect. The main pulse is split into two subpulses. In the first one, the polarization is rotated to $\theta_{<1,2>} = 30^\circ$ with respect to the second subpulse (main pulse), to conduct the NCS, and the second one is used to collide with the NCS $\gamma$-photons. $\ket{\parallel}$ and $\ket{\perp}$ denote photon polarization parallel and perpendicular to the first subpulse, respectively. The pair production rate $W_\pm(\ket{\parallel,\perp})$ is asymmetric (symmetric) for $\theta_{<1, 2>} = 30^\circ$ ($45^\circ$), and thereby with (without) the polarization purification, respectively.} \label{fig1}
	\end{figure}

\begin{figure}[!t]
		\includegraphics[width=\linewidth]{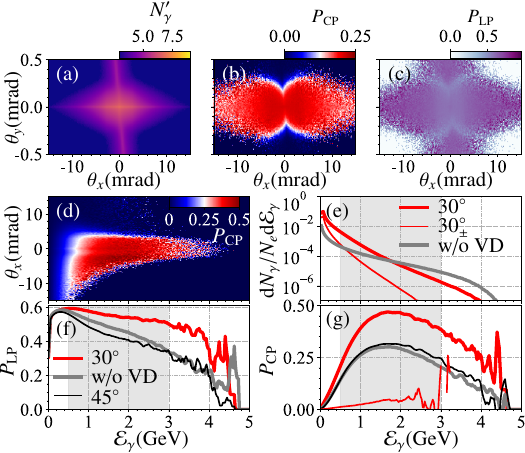}
		\caption{(a)-(c) Photon number density normalized to the electron number and scaled in logarithm $N'_\gamma \equiv \log_{10}[$d$^2(N_\gamma/N_e)/$d$\theta_x$d$\theta_y]$ (rad$^{-2}$), average circular polarization $P_{\rm CP} \equiv \overline{\xi}_2$ and linear polarization $P_{\rm LP} \equiv (\overline{\xi}^2_1 + \overline{\xi}_3^2)^{1/2}$ with respect to angles $\theta_x$ and $\theta_y$ \textcolor{black}{in the energy range of $500$-$3000$ MeV} (decorated as gray bands in (e)-(g), and $\theta_{x,y}\equiv \arctan(\frac{x,y}{z})$). (d)  $P_{\rm CP}$ with respect to the angle $\theta_x$ and energy $\mathcal{E}_\gamma$. (e) Photon energy spectra from primary electrons, $e^+e^-$ pairs, and without VD (pair production) are denoted as $30^\circ$, $30^\circ_\pm$,  and ``w/o VD'', respectively. (f) and (g) Energy-dependent $P_{\rm LP}$ and $P_{\rm CP}$ within $|\theta_x| \lesssim 5$ mrad and $|\theta_y| \lesssim 0.3$ mrad, respectively. Black lines label the results from the case of $\theta_{<1, 2>} = 45^\circ$, and other lines are the same denotations as those in (e). Due to poor statistics, the polarization line is broken for photons from the pair in the high-energy tail; see the thin-red line in (g).  } \label{fig2}
	\end{figure}

\begin{figure}
		\includegraphics[width=\linewidth]{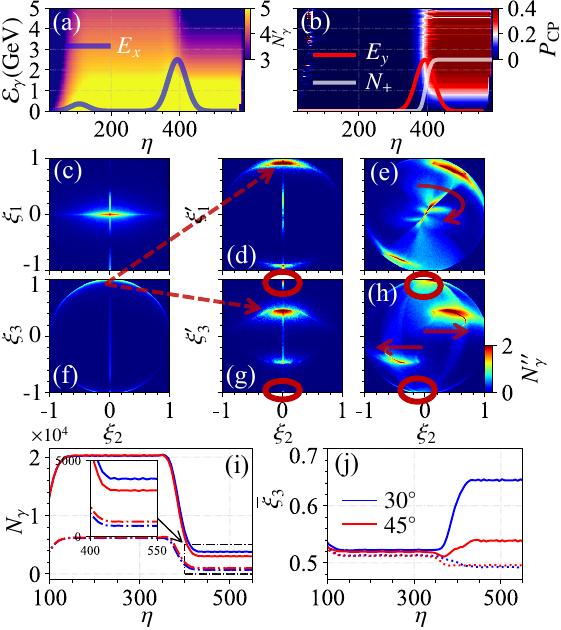}
		\caption{
			(a) and (b) Photon number density $N'_\gamma \equiv \log_{10}($d$^2N_\gamma/$d$\phi$d$\mathcal{E}_\gamma)$ (a. u.) and $P_{\rm CP}$ generated by the primaries. Solid lines indicate the relative temporal profile and $\eta \equiv \omega t - k z$ is the phase, respectively, of the laser.
			\textcolor{black}{Photon number distributions $N'' \equiv \log_{10}N_\gamma(\xi_2, \xi_{1,3})$ (a.u.) with respect to the Stokes parameters of the initial photons in (c) and (f), final photons without VB in (d) and (g), and final photons with VB in (e) and (h), respectively.} (i) Evolution of photon numbers in a sampled simulation for photons with $\mathcal{E}_\gamma > 1$ GeV, where blue and red lines show results for $\theta_{<1,2>}=30^\circ$ and $45^\circ$, and solid and dash-dotted lines represent photons with $\xi_3 > 0$ and $\xi_3 < 0$, respectively. 
			(j) Evolution of $\overline{\xi}_3$ corresponding to (i), with solid and dotted lines indicating cases of with and without considering the NBW process, respectively. Note that, in (i) and (j), $\xi_3$ is taken from the instantaneous frame. \textcolor{black}{Photons in the red circles in (g) and (h) are generated in the second subpulse. Red arrows in (e) and (h) indicate the VB effect. }} \label{fig3}
	\end{figure}

	\textcolor{black}{In this Letter, we put forward a novel method to generate high-brilliance and highly polarized $\gamma$-photon beams via the VD-assisted VB effect,} only utilizing a well-established unpolarized electron beam; see Fig.~\ref{fig1} for a schematic.
	In the conventional setup in Fig.~\ref{fig1}(a), an intense linearly polarized (LP) laser pulse colliding with a relativistic electron beam only generates brilliant LP  $\gamma$ photons via NCS. By contrast, in our proposed setup in Fig.~\ref{fig1}(b), an LP laser pulse is split into two subpulses: the first one is used to perform NCS and generate brilliant LP $\gamma$ photons;
the second one is used to partially convert the LP $\gamma$ photons into the CP ones via the VB effect (to ensure sufficient photon production in the first subpulse and significant VB effect in the second subpulse, the second subpulse is chosen to be relatively stronger than the first one). 
 Meanwhile, as shown in the dash-dotted box in Fig.~\ref{fig1}(b),  by fine-tuning the polarization angle $\theta_{<1,2>}$ between two subpulses, we find that NBW-induced VD effect can enhance the polarization of the $\gamma$-photon beam by changing the relative numbers of photons in different modes $\ket{\parallel}$ and $\ket{\perp}$, i.e., indirectly ``purifying" the beam polarization.
 The circular polarization degree can reach about $30\%(43\%)$ for energies above $500(1000)$ MeV and brilliance of about $10^{24}(10^{23})~\mathrm{photons / (s \cdot mm^2 \cdot mrad^{2} \cdot 0.1\%BW)}$ at $500~(1000)$ MeV; see details in Fig.~\ref{fig2}. And, it can be further improved up to 75\% by increasing the laser pulse duration.
Note that, the optimal $\theta_{<1,2>}$  is about $30^\circ$, not the $45^\circ$ used in the common VB detection methods \cite{Bragin_2017, King_2016_Vacuum, Nakamiya_2017}; see detailed explanations in Fig.~\ref{fig3}. Moreover, the circular polarization degree with respect to $\theta_{<1,2>}$ can serve as a reliable indicator of the VB effect.
In addition, our method is shown to be robust with respect to the laser and electron beam parameters; see Fig.~\ref{fig4} in the Appendix. 
	
	To simulate the interactions in Fig.~\ref{fig1}(b) self-consistently, the spin-resolved NCS and NBW are investigated by our Monte-Carlo methods~\cite{Li_2020_Polarized, Xue2020, Wan_2020, wan2023simulations}, incorporating with the VB effect by evolution of the Stokes parameters~\cite{Bragin_2017, wan2022enhanced, wan2023simulations}.
	Importance of the NCS, NBW, and VB effects is signified by the electron/photon nonlinear quantum parameter $\chi_{e,\gamma} = e\hbar/(m^2_ec^3) \sqrt{(F^{\mu \nu} p_\mu)^2} \simeq 2 \mathcal{E}_{e,\gamma}/(m_ec^2) a_0 (\hbar \omega_r/(m_ec^2)$, where $\hbar$, $-e$, $m_e$, and $c$ are the reduced Planck's constant, charge and mass of the electron, and the light speed in vacuum, respectively, $\mathcal{E}$ and $p_\mu$ are energy and four-momentum vector of the particle. $F^{\mu \nu}$,  $a_0 \equiv eE/(m_ec\omega_r)$, and $E$ are the field tensor, peak intensity, and electric field strength of the laser, respectively, and $\omega_r$ the reference frequency used for normalization.
	To describe the VB effect in a wide regime of $\chi_\gamma$, we use the accurate refractive indices described below.
	
	When an electron (positron) interacts with the laser field and generates a $\gamma$ photon via NCS, a set of Stokes parameters $\boldsymbol{\xi} = (\xi_1, \xi_2, \xi_3)$ with respect to polarization basis $\hat{e}_1$ and $\hat{e}_2$ are assigned to the photon \cite{berestetskii1982quantum, Baier_1998_Electromagnetic}.
	Here, $\hat{e}_1 \parallel \mathbf{E}_\mathrm{red., \perp}$, $\hat{e}_2 \parallel \hat{k} \times \hat{e}_1$, $\hat{k}$ is the direction of the photon momentum and $\mathbf{E}_\mathrm{red.} \equiv (\mathbf{E} + \hat{k} \times \mathbf{B})$ is the reduced electric field \cite{Toll1952}.
	$\xi_3$ indicates the linear polarization parallel to $\hat{e}_1$, $\xi_1$  linear polarization parallel to the direction with $\pi/4$ relative to $\hat{e}_1$, and $\xi_2$ the circular polarization \cite{McMaster1961}.
    In the subsequent evolution and pair production, the eigenbasis should be obtained with the local field.
    When the eigenbasis rotates by $\delta \psi$, the Stokes parameters transform via a rotation matrix 
    $
		\left(\begin{array}{l}
			\xi'_1 \\
			\xi'_3
		\end{array}\right) =
		\left(\begin{array}{ccc}
			\cos2\delta\psi & -\sin2\delta\psi \\
			\sin2\delta\psi & \cos2\delta\psi
		\end{array}\right)
		\left(\begin{array}{l}
			\xi_1 \\
			\xi_3
		\end{array}\right) $  \cite{Wistisen_2013_Vacuum, Bragin_2017}, where primed and unprimed Stokes parameters are evaluated in the updated and original frames, respectively. Note that primed  ones will denote Stokes parameters in the second subpulse in the latter.
	Before the photon decays into an $e^+e^-$ pair, its Stokes parameters evolve due to the VB effect. Within a single time step, the evolution can be represented by a phase rotation
	$
			\bigg(
			\begin{array}{c}
				\xi_1^f \\
				\xi_2^f \\
			\end{array}
			\bigg) = \bigg(
			\begin{array}{cc}
				\cos\delta\phi & -\sin\delta\phi \\
				\sin\delta\phi & \cos\delta\phi
			\end{array} \bigg) \bigg(
			\begin{array}{c}
				\xi_1^{i} \\
				\xi_2^{i}
			\end{array} \bigg)
		$~\cite{Wistisen_2013_Vacuum},
		where $i$ and $f$ indicate the initial and final Stokes parameters  of the $\gamma$ photon. Also, $\delta \phi= \int_0^{L} {\rm d}l (2\pi/\lambda_\gamma) \Delta n(\omega_\gamma)$ is the phase delay between two eigenstates for a  photon (wavelength $\lambda_\gamma$  and frequency $\omega_\gamma$) traversing a polarized vacuum of length $L$, 
		and $\Delta n = n_\perp - n_\parallel$,  with $\parallel$ and $\perp$ denoting the polarization modes parallel to $\hat{e}_1$ and $\hat{e}_2$, respectively.
		$n_{\parallel,\perp}$ follow from the real and imaginary parts
		$\mathrm{Re}(n) = 1 - \frac{45}{4} \mathcal{D} \int_0^1 d\upsilon f(\upsilon) \left[ \pi x^{4/3} \mathrm{Gi}'(x^{2/3}) \right]$ and 
   $\mathrm{Im}(n) = \frac{45}{4} \mathcal{D} \int_0^1 d\upsilon  f(\upsilon) \left[\frac{x^2}{\sqrt{3}}\mathrm{K}_{2/3} \left( \frac{2}{3}x \right) \right]$ of the refractive index $n$ \cite{Shore_2007_Superluminality, Bragin_2017,wan2022enhanced}. Here,  
		$\mathcal{D} = \frac{\alpha}{90\pi} \left(\frac{e|{\bf E}_{\rm red,\perp}|}{m_e^2}\right)^2 \equiv \frac{\alpha}{90\pi} \frac{\chi_\gamma^2}{\omega^2/m_e^2}$, 
		$f(\upsilon) = (1-\upsilon^2) \left\{ \begin{matrix} \frac{1}{2}(1 + \frac{1}{3}\upsilon^2) \\ 1 - \frac{1}{3}\upsilon^2 \end{matrix}  \right\}$, 
		with $\mathrm{Gi}'(x)$ being the derivative of the Scorer function, $\mathrm{K}_{\nu}(x)$ the modified Bessel function of the second kind, and $x = 4/(1-\upsilon^2)\chi_\gamma$.
	The first and second rows of $f(\upsilon)$ denote the polarization states parallel to $\hat{e}_1$ and $\hat{e}_2$, respectively, i.e., labeled with $\parallel$ and $\perp$.
		Re($n$) accounts for the photon propagation, and Im($n$) accounts for the $e^+e^-$ pair production, which is resolved by the NBW process \cite{Toll1952, Shore_2007_Superluminality, Borysov2022}.
	In each time step for $\gamma$-photon propagation, the Stokes parameters will first be mapped to the instantaneous frame of $\hat{e}_1$, $\hat{e}_2$, and the NBW is checked.
	If no pairs are created, the polarization may change due to the selection effect \cite{yokoya2011cain}, and then the photon will continue to propagate including the VB-induced phase delay, i.e., rotating the Stokes parameters.
	
 As an example, the total intensity of the laser is $I_0 \simeq 5.48\times10^{22}~\mathrm{W/cm^2}$ (corresponding to $a_0 \simeq 200$), pulse duration $\tau = 10T_0$ ($T_0 = \lambda /c$), focal radius $w = 3~\mathrm{\mu m}$ and wavelength $\lambda = 1~\mathrm{\mu m}$. The first subpulse is $a_1 = 30$ and the second one with $a_2 = \sqrt{a_0^2 - a_1^2} \simeq 197.7$. 
 \textcolor{black}{The main pulse is split into two subpulses by using a beamsplitter before the compressor, and then two subpulses can be injected into two independent compressors \cite{Cole2018, Poder2018, PhysRevLett.113.224801, Gales_2018_extreme, Turner2022, Yan2017}. As two subpulses share the same oscillator and amplifier, it can avoid the problem of jitter in the time synchronization \cite{PhysRevLett.113.224801}. Moreover, the synchronization of two subpulses can be further optimized by using the optical synchronization techniques \cite{Corvan:16, Poder2018, 10.1063/5.0115918}.}
	The polarization of the first subpulse is then rotated to $30^\circ$ with respect to the original polarization (the second one). 
	We use a feasible electron beam with a peak energy of 5~GeV, energy spread of $3\%$, angular spread of $0.1~\mathrm{mrad}$, longitudinal size of $5~\mathrm{\mu m}$, transverse radius of $0.4~\mathrm{\mu m}$, and number of $6.25\times 10^7$ (total charge of 10 pC) \cite{Esarey2009, Gonsalves2019, Aniculaesei2023}.
 In the absence of the dense LSP electron beams from LWFA in experiments, here we employ the well-generated unpolarized electron beams.
	
In all simulations, the first subpulse is assumed to be polarized along the $x$-axis. The maximum photon yield per electron can reach $10^{8}/\mathrm{rad}^2$; see
	Fig.~\ref{fig2}(a). The peak degrees of circular and linear polarization can reach about $45\%$ and $73\%$, respectively; see Figs.~\ref{fig2} (b)-(d). \textcolor{black}{See the full energy range version in Supplementary Material (SM) \cite{SM}.}
	In certain applications (e.g., photoproduction of $\pi^0$ pairs from nucleons \cite{PhysRevLett.125.062001} and deuteron photodisintegration \cite{BASHKANOV2023138080}), one can obtain about $8\times 10^6$ ($4\times 10^7$) $\gamma$ photons per shot with $\mathcal{E}_\gamma \gtrsim 1000$ MeV (500 MeV) and a circular polarization degree of approximately $43\%$ ($30\%$) [Figs.~\ref{fig2}(e), (g) and Fig.~\ref{fig4}].
 \textcolor{black}{The circular and linear polarization of the photons is unlikely to be separated via the angle filtering as the circular one originates from the linear one via the VB effect; see Figs.~\ref{fig2}(b) and (c). Fortunately, when the laser pulse duration increases properly, the circular polarization can be enhanced, and even higher than the linear one; see Fig.~\ref{fig4}(e). Moreover, when $\theta_{<1,2>}\simeq n\pi$ ($n$ is an integer), the beam is highly LP with a degree of about 80\%, which is much higher than directly obtained via the NCS \cite{Li_2020_Polarized}; see Fig.~\ref{fig4} (f).
Note that the VD effect will consume high-energy photons, which decay into $e^+e^-$ pairs, and these pairs can further emit low-energy photons, while both the circular and linear polarization can be enhanced; see Figs.~\ref{fig2}(e), (f) and (g).}
  In addition, the circular polarization is mainly contributed by photons generated by primary electrons in the first subpulse; see Fig.~\ref{fig2}(g).
	Secondary photons from NBW $e^+e^-$ pairs generated in the second subpulse, can barely be CP (with a degree of about $1\%-2\%$).

As shown schematically in Fig.~1 (b), the generation of high-energy CP $\gamma$ photons is accomplished in two successive stages: NCS in the first subpulse, followed by VB (accompanied by NBW and NCS cascade) in the second subpulse. In the first stage, only primary electrons emit photons, and the NBW pair production is negligible due to the low intensity of the first subpulse; see the evolution of photon spectra in Fig.~\ref{fig3}(a) and normalized positron number $N_+$ in Fig.~\ref{fig3}(b).
	Since radiation loss by an electron is proportional to the laser energy via $\delta \gamma_e \propto a^2 \gamma_e^2 \omega^2 \tau_L$ \cite{Thomas2012}, with a stronger scattering laser more photons are emitted by primary electrons in the second stage; see Fig.~\ref{fig3}(a), but with much lower energies than in the first stage.
	In the second stage, high-energy $\gamma$ photons can create $e^+e^-$  pairs via the NBW process; see Fig.~\ref{fig3}(b).
	These pairs, in turn, can emit abundant photons which is negligible for the final photon energy spectrum and polarization in the energy range of $\mathcal{E}_\gamma > 500$ MeV; see Figs.~\ref{fig2}(e) and (g). Besides, further cascade is suppressed  due to the decrease of $\chi$.

All these photons are initially LP along the external field (either in the first or second subpulse) with an average Stokes parameter of $\xi_3 \simeq 0.6$ ($\mathcal{E}_\gamma / \mathcal{E}_e \simeq 0$) and then decreasing almost linearly to 0 for $\mathcal{E}_\gamma / \mathcal{E}_e \simeq 1$ \cite{Li_2020_Polarized}. In the whole energy range, one has $\overline{\xi}_1 \simeq \overline{\xi}_2 \simeq 0$; see Fig.~\ref{fig3}(c).
	For photons emitted in the first stage, there is an angle $\delta \psi = \theta_{<1, 2>}$ between the instantaneous eigenbasis and the polarization of the second subpulse. Subsequently, due to the change of the instantaneous eigenbasis, their Stokes parameters change as well when they enter the second subpulse; \textcolor{black}{see Figs.~\ref{fig3}(d) and (g), where photons are redistributed from $\xi_3 \simeq \pm 1$ to $\xi'_1 \simeq \pm \sin2\theta_{<1,2>} \simeq \pm \sqrt{3}/2$ and $\xi'_3 \simeq \pm \cos2\theta_{<1,2>} \simeq \pm 1/2$; see the arrow indication from Fig.~\ref{fig3}(f) to Figs.~\ref{fig3}(d) and (g) for the case of $\xi_3 \simeq +1$.
 In principle, to obtain high circular polarization via the VB effect, $\xi'_1$ should acquire a maximum with $\theta_{<1, 2>} = \pm \pi / 4$ ($45^\circ$), and $\xi'_3 \simeq 0$.}
	When these photons propagate in the second subpulse, some of them decay into $e^+e^-$ pairs, and others experience the VB effect, i.e., $\xi_1'$ transfers to $\xi_2$; \textcolor{black}{see the rotation and translation in Figs.~\ref{fig3}(e) and (h)}.
	However, for photons emitted in the second subpulse, the polarization basis at the instant of creation, and during propagation or pair-production, is the same ($\hat{e}_1 \parallel \mathbf{E}_\mathrm{red., \perp}$ and $\delta \psi = 0$).
	Thus, for these photons, $\overline{\xi'}_1 \simeq 0$, and the VB effect is negligible (VB only transfers polarization between $\xi'_1$ and $\xi_2$); \textcolor{black}{see these photons in the red circles of Figs.~\ref{fig3}(g) and (h).}
	Besides, since the energy of photons generated in the second subpulse is relatively smaller than that of those generated in the first subpulse, they affect the final circular polarization degree negligibly; see Fig.~\ref{fig3}(a). Moreover, the deflection of the primary electrons before scattering with the second subpulse does not significantly affect the final circular polarization; see Fig. S4 in \cite{SM}.

	An interesting finding that sounds counterintuitive is the optimal angle between the two subpulses is not $45^\circ$ which is usually used in the common VB detection methods \cite{Bragin_2017, King_2016_Vacuum, Nakamiya_2017}.
	After embedding the QED cascade of NCS and NBW, the simulations show that the case of $\theta_{<1, 2>} = 30^\circ$ can yield a much higher circular polarization degree than that in the case of $45^\circ$; see Fig.~\ref{fig2}(g).
	This contradiction arises when the polarization-dependent NBW acts as a purifier of the photon polarization, i.e., the VD effect. The circular (linear) polarization degree of NCS photons could increase by over 30\% (20\%); see Figs.~\ref{fig2}(f) and (g). 
 This increment in the circular polarization degree is not an expected result in the common VB or VD detection proposals, where an exponential decay $e^{-(\lambda_1 + \lambda_2)}$ is presented before all Stokes parameters due to the VD effect ($\lambda_{1,2}>0$ are related to the pair production rate) \cite{Bragin_2017}.
As indicated in Fig.~\ref{fig1}(b), this VD-assisted purifying mechanism stems from the NBW polarization-dependent pair-production rate, which has the form of $W_\pm \propto W_0 + \xi'_3 W_3$, with $W_0$ only depending on the photon energy and $W_3$ being the polarization-dependent term \cite{Baier_1998_Electromagnetic}. 
	For photons generated by NCS, the average polarization is $\bar{\xi_3} \simeq 60\%$ for the low energy range and $\simeq 0$ for the high energy range. 
	This means that for low-energy photons, about $20\%$ and $80\%$ of photons acquire $\xi_3 = -1$ and $1$, and for high-energy photons, about $50\%$ and $50\%$, respectively.
	When these photons enter the second subpulse, the Stokes parameters get rotated into the new frame with $\xi'_3 = \cos2\theta_{<1,2>} \xi_3$ and $\xi'_1 = \sin2\theta_{<1, 2>}\xi_3$.
	Therefore, for $\theta_{<1, 2>} = 45^\circ$ ($\bar{\xi'_3} \simeq 0$), the pair-production rate is symmetric for $\xi'_1 = \pm \sin2\theta_{<1,2>}\simeq \pm 1$. 
	However, for $\theta_{<1, 2>} \neq 45^\circ$ ($\bar{\xi'_3} \neq 0$), it is no longer symmetric for $\xi'_1 \simeq \pm \sin2\theta_{<1, 2>}$; see Fig.~\ref{fig3}(i).
	More photons with $\xi_3 > 0$ and fewer ones with $\xi_3 < 0$ are left for the $30^\circ$ than $45^\circ$. 
	This asymmetric pair-production rate results in the increase of $\bar{\xi'_1}$ and, therefore, can yield higher $\bar{\xi_2}$; see the final $\bar{\xi_3}$ ($\simeq \bar{\xi'_1}/\sin2\theta_{<1,2>}$)  in Fig.~\ref{fig3}(j).  
	In principle, this angle can be optimized via the $n$-step pair-production rate
	$
		\max(\bar{\xi_1'}) = \max\left(\xi_1'\frac{\mathcal{N}_+ - \mathcal{N}_-}{\mathcal{N}_+ + \mathcal{N}_-}\right) = \max\left(\sin2\theta_{<1, 2>}\frac{\mathcal{N}_+ - \mathcal{N}_-}{\mathcal{N}_+ + \mathcal{N}_-}\right)
	$,
	where $N_\pm$ and $\mathcal{N}_\pm \equiv N_\pm [1-(W_0 - \xi'_{3,\pm} W_3)]^n = N_\pm [1-(W_0 \mp \cos2\theta_{<1,2>}W_3])]$ denote the total and non-decay photon numbers with $\xi_{3, \pm} = \pm 1$, respectively.
	With current laser and electron parameters, owing to the VD enhancement to the polarization of NCS photons, we obtain the optimal $\theta_{<1, 2>} \simeq 30^\circ$, not the $45^\circ$ used in the common VB detection methods \cite{Bragin_2017, King_2016_Vacuum, Nakamiya_2017}.
	
For experimental feasibility, the impact of the laser parameters on the photon yield and polarization has been investigated, as shown in the Appendix, and the results show that our method is robust with respect to the laser intensity, pulse duration, and relative polarization angle of two subpulses. 
	In particular, with a longer pulse duration of $\tau \simeq 25T_0$, the average circular polarization can reach about  60\% and 75\% for energies over 500 MeV and 1000 MeV, respectively. Besides, the VD effect can induce anomalous polarization, which can also be used for the confirmation of the VD effect itself.
	
	In conclusion, we put forward a novel method for generating high-brilliance \textcolor{black}{LP and CP $\gamma$-photon beams via the VD-assisted VB effect}, which has significant applications in material physics, nuclear physics,  astrophysics, high-energy particle physics, new physics beyond the Standard Model, etc. And, another particularly intriguing outcome of our method is the potential confirmation of  the well-known VB effect itself, which was predicted more than eighty years ago but has not been directly observed in experiments yet.
	
	\vskip 0.2cm
	{\bf Acknowledgments:} This work is supported by the
	National Key Research and Development (R\&D) Program under Grant No. 2022YFA1602403, the National Natural Science Foundation of China (Grants No. U2267204, No. 12022506, No. 12005305, No. 12275209, and No. 12105217), the Foundation of Science and Technology on Plasma Physics Laboratory (No. JCKYS2021212008), the Shaanxi Fundamental Science Research Project for Mathematics and Physics (Grant No. 22JSY014), the Fundamental Research Funds for Central Universities (No. xzy012023046), and the Foundation under Grants No. FY222506000201 and No. FC232412000201.
	
		\begin{figure}[!t]
			\includegraphics[width=\linewidth]{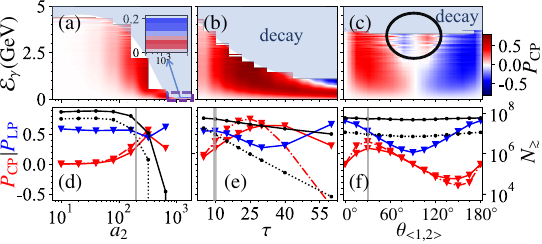}
			\caption{Impact of the laser parameters on the VB effect and polarization degree of the generated $\gamma$ photon with respect to 
				[(a) and (d)] second subpulse intensity $a_2$, [(b) and (e)] pulse duration $\tau$,
				[(c) and (f)] relative polarization angle $\theta_{<1, 2>}$. 
				In the first column, the light-blue area with ``decay'' indicates no photons can be detected (due to pair production).
				The black circle in (c) denotes the anomalous polarization region.
				In the second column, solid and dash-dotted lines denote the results of photons with energies $\mathcal{E}_\gamma \gtrsim 500$ MeV and $1000$ MeV, respectively. All results are collected within an angle of $|\theta_{x, y}| \leq 5$ mrad. \textcolor{black}{Red, blue, and black lines indicate $P_{\rm CP}$, $P_{\rm LP}$, and the total number $N_\gtrsim$.}
				The vertical gray lines indicate parameters used in Figs.~\ref{fig2} and \ref{fig3}. All other parameters are the same as those in Fig.~\ref{fig2}.
			} \label{fig4}
		\end{figure}
		
			\vskip 0.2cm
			{\bf Appendix on the experimental feasibility.} To demonstrate the experimental feasibility of our method, we focus on the dimensionless intensities $a_{1,2}$ of the two subpulses, pulse duration $\tau$, and their relative polarization angle $\theta_{<1,2>}$.
		For a specific laser facility, the total intensity is fixed, e.g., $a_0 \simeq 200$. By varying the relative intensity of both subpulses, the yield number is quite stable, but the circular polarization degree decreases for $a_1 \gtrsim 100$; see in \cite{SM}. 
		For a fixed $a_1$ and a varied $a_2$ over a wide range, the circular polarization degree continuously increases until $a_2 \simeq 400$ and more photons are converted into pairs.
		Increasing $a_2$ further induces the VB phase retardation $\delta \phi \gtrsim \pi$ for high-energy photons, and one has $\xi_2 \simeq \sin \delta \phi \xi_1' < 0$; see Fig.~\ref{fig4}(a).
		\textcolor{black}{This pattern also applies to the pulse duration $\tau$. In the second subpulse, the linear polarization can be transferred to the circular one, i.e., $\xi'_1$ is transferred to $\xi_2$ due to the VB effect via $\xi_2 \propto \xi'_1 \sin\delta\phi$, with $\delta \phi \propto \frac{\int_0^L 2\pi \Delta n(\omega)}{\lambda_\gamma} dl \propto \frac{\Delta n(\omega_\gamma) \tau}{\lambda_\gamma}$, and $L \propto c\tau$ in the external field. For $\delta \phi \ll 1$, one has $\xi_2 \propto \xi'_1\delta\phi \propto \xi'_1 \tau$, i.e., the circular polarization is proportional to the pulse duration of the second subpulse; see cases with $\tau \lesssim 20T_0$ in Fig.~\ref{fig4}(e). While for much longer pulse durations, on the one hand, $\xi_2 \propto \xi'_1 \sin\delta\phi$ is periodic with respect to the phase retardation $\delta\phi$. On the other hand, more photons will decay into pairs. These two effects will both decrease the circular polarization degree; see cases of $\tau \gtrsim 30T_0$ in Fig.~\ref{fig4}(e).} And high-energy photons acquire negative $\xi_2$ for larger $\tau$.
		The optimal pulse duration of the currently available laser facility is in the range of $10 T_0-30 T_0$. 
		In the case of $\tau = 25 T_0$, the circular polarization degree can reach $60\%$ and $75\%$ for photon energy over 500 MeV and 1000 MeV, respectively.
		Relative polarization angle $\theta_{<1,2>}$ of the two subpulses is another key parameter for generating CP $\gamma$ photons via the VB effect.
		As stated above, for photons emitted in the first stage, $\xi_3$ at creation is first mapped into $\xi'_1$ via $\theta_{<1, 2>}$, and then rotated to $\xi_2$ via the VB effect in the second stage.
		This mapping is periodic with respect to $\theta_{<1, 2>}$, and so is the induced circular polarization degree; see Figs.~\ref{fig4}(c) and (f).
		For photons with $\mathcal{E}_\gamma < 500$ MeV, the VD purifying mechanism is relatively weak, and the circular polarization degree reaches 30\% for $\theta_{<1,2>} \in [30^\circ, 45^\circ]$ or $[135^\circ, 150^\circ]$. 
		For photons with $\mathcal{E}_\gamma \gtrsim 1$ GeV, the purifying mechanism amplifies the circular polarization degree at $\theta_{<1,2>} \simeq 30^\circ$ or $150^\circ$, beyond the results corresponding to $\theta_{<1, 2>} \simeq 45^\circ$ or $135^\circ$.
		However, as shown in Fig.~\ref{fig4}(c), an anomalous polarization regime (within the black circle) exhibits a circular polarization degree opposite that of its neighboring region.
		This anomalous CP reversal also originates from the NBW process.
		For $\theta_{<1,2>} > 45^\circ$, $\overline{\xi'}_3 \simeq \cos 2\theta_{<1,2>} \overline{\xi}_3 \lesssim 0$, which indicates that photons with $\xi_3 = 1$ exhibit larger pair-production rates than that with $\xi_3 = -1$.  
		Although initially $N_\gamma(\xi_3 = 1) > N_\gamma(\xi_3 = -1)$, this inequality becomes weak with the increase in photon energy and may reverse due to pair production.
		For $N_\gamma(\xi_3 = 1) < N_\gamma(\xi_3 = -1)$, $\overline{\xi'}_1 \simeq \overline{\xi}_3 \sin 2 \theta_{<1, 2>} \propto [N_\gamma(\xi_3 = 1) - N_\gamma(\xi_3 = -1)] \sin2\theta_{<1, 2>}<0$, i.e., $\overline{\xi}_2$ will reverse sign; see the region of $\mathcal{E}_\gamma \in [2, 3.5]$ GeV and $\theta_{<1,2>} \in [45^\circ, 90^\circ]$ in Fig.~\ref{fig4}(c).
		The same analysis applies to the region of $\mathcal{E}_\gamma \in [2, 3.5]$ GeV and $\theta_{<1,2>} \in [90^\circ, 135^\circ]$.
		Besides, since the total laser intensity is the same, the number of generated photons is relatively stable.
  \textcolor{black}{Moreover, as mentioned previously, when $\theta_{<1,2>} \simeq 0^\circ$, the linear polarization can reach the maximum to about 80\%; see Fig.~4(f). This is due to that the linear polarization can also be enhanced by the VD effect; see more details in \cite{SM}. As the parameters are further optimized, the linear polarization degree can be even higher.}
  Therefore, this scheme is robust with respect to laser and electron parameters. \textcolor{black}{In addition, since both the laser and electron beam parameters can affect the yields and polarization of the photon, a tradeoff should be done to fulfill the requirements of some specific applications.}

\bibliography{vblib}
	
\end{document}